# COVID-19 pandemic: a mobility-dependent SEIR model with undetected cases in Italy, Europe, and US

Pandemia di COVID-19: un modello SEIR dipendente dalla mobilità e con stima dei casi nascosti in Italia, Europa e Stati Uniti

Nicola Picchiotti,[1,2*°] Monica Salvioli,[3] Elena Zanardini,[4] Francesco Missale[5,6*]

[1] Department of Mathematics, University of Pavia (Italy)
[2] Internal Model Validation, Banco BPM spa, Verona (Italy)
[3] Department of Mathematics, University of Trento (Italy)
[4] Department of Medical and Surgical Specialties, Radiological Sciences and Public Health, University of Brescia (Italy)
[5] Department of Molecular and Translational Medicine, University of Brescia (Italy)
[6] IRCCS Ospedale Policlinico San Martino, Genova (Italy)

* These authors equally contributed to the work

° The views, thoughts and opinions expressed in this report are those of the authors in their individual capacity and should not be attributed to Banco BPM or to the authors as representatives or employees of Banco BPM

Corresponding author: Monica Salvioli; monica.salvioli@polimi.it

## ABSTRACT

**OBJECTIVES:** to describe the first wave of the COVID-19 pandemic with a focus on undetected cases and to evaluate different post-lockdown scenarios.
**DESIGN:** the study introduces a SEIR compartmental model, taking into account the region-specific fraction of undetected cases, the effects of mobility restrictions, and the personal protective measures adopted, such as wearing a mask and washing hands frequently.
**SETTING AND PARTICIPANTS:** the model is experimentally validated with data of all the Italian regions, some European countries, and the US.
**MAIN OUTCOME MEASURES:** the accuracy of the model results is measured through the mean absolute percentage error (MAPE) and Lewis criteria; fitting parameters are in good agreement with previous literature.
**RESULTS:** the epidemic curves for different countries and the amount of undetected and asymptomatic cases are estimated, which are likely to represent the main source of infections in the near future. The model is applied to the Hubei case study, which is the first place to relax mobility restrictions. Results show different possible scenarios. Mobility and the adoption of personal protective measures greatly influence the dynamics of the infection, determining either a huge and rapid secondary epidemic peak or a more delayed and manageable one.
**CONCLUSIONS:** mathematical models can provide useful insights for healthcare decision makers to determine the best strategy in case of future outbreaks.

**WHAT IS ALREADY KNOWN**
■ The outbreak of Coronavirus disease 2019 (COVID-19) emerged in China at the end of 2019 has spread worldwide within few weeks, reaching a pandemic status.
■ China and Italy, followed by other countries, have implemented control measures, varying in level of effectiveness and strength, which were proven to be successful.

**WHAT THIS STUDY ADDS**
■ The challenge is to find the best strategy to progressively relax the control measures keeping the number of new infections below a certain threshold.
■ This mathematical model can be used to predict the evolution of the epidemic, and in particular the number of undetected cases, under different health policies. It can be easily recalibrated and applied in case of future outbreaks.

## RIASSUNTO

**OBIETTIVI:** descrivere la prima ondata dell'epidemia di COVID-19, con particolare attenzione ai casi nascosti e valutare diversi scenari *post-lockdown*.
**DISEGNO:** lo studio introduce un modello SEIR a compartimenti, che tiene conto dei casi nascosti nelle diverse regioni, degli effetti delle restrizioni sulla mobilità e delle misure di contenimento della diffusione del contagio adottate, come l'utilizzo della mascherina e il lavaggio frequente delle mani.
**SETTING E PARTECIPANTI:** il modello è validato utilizzando i dati delle regioni italiane, di alcuni Paesi europei e degli Stati Uniti.
**PRINCIPALI MISURE DI OUTCOME:** l'accuratezza delle previsioni del modello è misurata attraverso l'errore medio assoluto percentuale (MAPE) e i criteri di Lewis. I parametri sono in accordo con quelli riportati nella recente letteratura.
**RISULTATI:** il modello fornisce la stima delle curve epidemiche in diversi Stati e il numero di casi nascosti e asintomatici, che rappresenta la più probabile futura fonte di infezione. Il modello è stato applicato a un caso di studio relativo alla provincia cinese di Hubei, che per prima ha allentato le restrizioni sulla mobilità. I risultati mostrano che sono possibili diversi scenari. La mobilità e l'utilizzo di misure di protezione individuale influenzano notevolmente la dinamica dell'infezione e possono portare a un secondo picco improvviso e pericoloso oppure a uno più lontano nel tempo e più gestibile.
**CONCLUSIONI:** i modelli matematici come quello introdotto possono rappresentare uno strumento utile nei processi decisionali in sanità pubblica, per determinare la strategia migliore nel caso di future epidemie.

Parole chiave: COVID-19, modelli matematici, SEIR, sanità pubblica, epidemiologia, *lockdown*

## INTRODUCTION AND OBJECTIVES OF THE STUDY

Italy was the first European country experiencing a substantial outbreak of the Coronavirus disease 2019 (COVID-19) and consequently applying lockdown emergency measures to contain the spread of the epidemic, similarly to the Chinese strategy. In particular, as the number of infections was growing exponentially,[1] the Italian authorities decided to close all schools on 05.03.2020 and introduced community containment lockdown measures starting on 9 March. Spain, France, Germany, United Kingdom (UK), and United States of America (USA) are the countries with the highest number of infections and deaths. All these countries adopted public health measures and, most of them, implemented a complete lockdown, but such measures were introduced at different time points following local outbreaks and with different strengths.[1]

The aim of this work is to model the evolution of the COVID-19 pandemic in different regions, taking into account major determinants such as undetected cases, related to the different case fatality ratio measured and taking into account background death data, and different mobility changes as effect of public health measures.

Assuming that isolation is successfully applied to the positive detected cases, undetected and asymptomatic cases will represent the main source of infections in the near future, potentially determining new outbreaks. This poses a crucial challenge for public health decision makers, namely to determine when and how to relax the control measures implemented so far, taking into account the whole number of estimated infectious individuals and not only the tested cases. This model provides an accurate estimation of the undetected and asymptomatic cases, therefore allowing to predict future epidemic peaks and design the most effective way to relax the control measures, in terms of timing and strength in term of mobility restrictions and adoption of personal protective behaviour. Moreover, it can be expanded and tailored when more and new region-specific data will become available.

## THE MOBILITY-DEPENDENT SEIR MODEL

This study models the spread of the COVID-19 pandemic in different geographical regions using a modified SEIR compartmental model,[2] fitted using publicly available data from the 2019-nCoV data repository by Johns Hopkins CSSE and from the Italian official repository updated as of 24 April.[3]

In a classical SEIR model the population is divided into four classes: susceptible, exposed, infected, and recovered. In this specific case, illustrated in on-line supplementary materials, the classes of positive tested, undetected infected, and – both for detected and undetected ones – the two possible ending point, healing and death (figure 1A), were added. Disease-specific parameters were derived from the recent literature, including a mean incubation time ($\eta^{-1}$) of 5.1 days,[4] a mean time to isolation ($\gamma_1^{-1}$) of 4.8 days,[5] a mean recovery time for undetected cases of 10 days,[6] and time to death from onset of 17.8 days.[7] The time of recovery notification by the public health system after isolation of the detected cases was fitted. For the lethality rate of COVID-19, the time-delay adjusted case fatality rate (aCFR) of 1.4% (CI 1.2%-1.7%) derived by Shim et al.[8] was used. Available Italian death records data from years 2015-2020 (considering only February and March 2020) were compared and it was assumed that the increase in deaths registered in 2020 until 28 March in each Italian Region is entirely related to the COVID-19 pandemic, even though only a fraction was confirmed as positive. In particular, it was estimated that undetected COVID-19 cases in Italy represent 53% of the registered COVID-19 deaths (table 1), which is consistent with other scenarios. For instance, on 17 April 2020 the Chinese government announced that in Hubei Province there were 1,290 COVID-19-related deaths (occurred from the beginning of the pandemic) which were not previously registered as COVID-19 deaths and which represented about 40% of the officially registered deaths.[9] This fact supports the use of the Italian estimation for all analysed countries.

Therefore, by combining the background death records data with the measured naïve case fatality rates (nCFRs) up to 10 March, ranging from 2.26% in Germany to 18.27% in Lombardy Region (Northern Italy), it is possible to give a reasonable estimate of the proportion of undetected cases in different Regions to obtain a target adjusted case fatality rate (aCFR) of 1.4% (95%CI 1.2-1.7%).[8]

Public health measures were introduced in the model, taking into account the effects of mobility changes as well as the potential effects of adopting a personal protective behaviour, like frequent hands washing and wearing a mask. The former is modelled with a decreasing logistic function, where the target plateau is the average mobility change reported by Google COVID-19 community mobility report in each geographical area over the last 7 available observations (11-17 April), compared to the average value measured between 3 January and 6 February 2020. Each community mobility report displays the change in visits to places like grocery stores, retails, transit stations, working and residential places, and parks; reports are created with aggregated, anonymised sets of data from users who have turned on the location history (a Google account-level setting) on mobile devices. The shape of the logistic function is calibrated knowing the date of the implementation of public health measures (physical distancing advisement or school closures)[1] and date of the maximum reduction of the mobility (figure 1 B to D). The effect of the adoption of a personal protective behaviour as frequent handwashing or masks wearing, recommended worldwide,[10] was derived from a metanalysis of studies performed in the context of SARS epidemic,[11] estimating an odds ratio (OR) of the use of masks of 0.32 (95%CI 0.25-0.40) and an OR of frequent handwashing of 0.45 (95%CI 0.36-0.57) for the risk of infection. The models are fitted until 17 April leaving the last 7 observed days for the accuracy analysis of model predictions through mean absolute percentage error (MAPE) analysis.[12] The model is deterministic, thus the reasonable confidence intervals of fitted estimates were derived considering the 95%CI of the aCFR and the 95%CI of personal protective measures ORs derived from literature data.[8,11] Mathematical details of the model, including equations, parameters calibration, and fitting procedure are illustrated in the supplementary materials.

## RESULTS

### PANDEMIC DYNAMICS IN DIFFERENT GEOGRAPHICAL CONTEXTS

In the Italian scenario, the undetected proportion of cases ranges from 76% in Lazio to 96% in Sardinia, being 93% for the whole Italy (table 1, figure 2A). Epidemic curves obtained using the calibrated SEIR models are reported in figure 2B; the accuracy analysis of forecasting showed a median MAPE of 9.2%; interquartile range (IQR) 3.6%-14.5% with a range of 0.9%-34.8% (table 1). The forecasting procedure permits an estimate on 24 April of the whole fraction of infected population, including undetected cases, ranging from 0.4% (95%CI 0.3-0.5) for Campania Region (Southern Italy) to 13.6% (95%CI 13.4-13.9) for Lombardy, being ≈ 5 to 40-fold higher than positive tested cases, as shown in table 1 and in heatmaps in figure 2A. This is coherent with recent literature results estimating the COVID-19 prevalence, until 7 April, ranging from 0.35% to 11.2% across different Italian Regions.[13]

In the international context, the number of supposed undetected cases ranges from 60% in Germany to 92% in UK. The application of the SEIR models, tested in the whole Italy, France, Germany, UK, Spain, and USA, was satisfactory, permitting also to show the trajectories of overall cases, including undetected ones (figure 3A). Analysing the forecasting accuracy in the international context, the median MAPE was 21.9% (IQR 12.3-28.8; range 8.5-31.8%) (table 1), statistically comparable to that of Italian Regions (p=0.08) by Mann-Whitney test. Considering the Lewis criteria[12] for the forecasting accuracy judgement, the model here presented was highly accurate in 12 cases (46%), good in 7 (27%), and reasonable in 7 (27%).

The best fit in each Region of the recovery notification time ($1/\gamma_2$), has median value of 63 days (IQR 38-99; range 15-10,000 days), possibly representing different efficiency of recovery reporting in each geographical scenario. The estimation of the basic reproductive number ($R_0$) was derived from the model through the Next generation matrix method.[14] The contribution of positive detected cases or undetected ones for the generation of secondary cases is weighted to obtain the overall reproductive number ($R_0$), representing the overall number of newly cases generated along the infectious time of an affected case (table 1). The mean $R_0$ related to detected cases ($R_{0P}$) is 7.5±2.1, the one related to undetected ones ($R_{0H}$) 15.7±4.5, determining an overall weighted $R_0$ value of 15.1±4.3, in good agreement with models using different approaches.[15]

### MODELLING THE LOCKDOWN RELAXATION

Knowing the daily fraction of undetected infectious cases on overall infectious ones (average 92.0±5.6%), it was possible to estimate that 91.8% of new positive detected cases were determined by undetected ones until 24 April; this explaining the 91.4% of the total new infections (table 1). Specifically, the results in the Chinese scenario (86.7% and 85.2%, respectively) are in agreement with previous literature.[16] Therefore, the possibility to estimate the amount and the dynamics of undetected cases is a key point when planning to loosen lockdown measures.

The model was applied for the Hubei case study, where on 7 April the Chinese government decided to relax lockdown measures, showing different possible scenarios (figure 3B). According to the model, there were 2,502 (CI95% 2,251-3,205) infectious subjects on 7 April, most undetected (96.4%). It was shown how the infection dynamics depend on the level of target mobility that is restored, compared to the baseline, and on the adoption of personal protective measures. These two factors can make the difference between a huge and rapid secondary epidemic peak or a more delayed and manageable one (figure 3B).

## DISCUSSION AND PERSPECTIVES

Despite the proven effectiveness of lockdown measures, as shown in the Chinese epidemic, the challenge is now to assess their efficacy and to plan strategies to progressively relax them.[17,18] Mathematical models can help determine the trajectories of the pandemic, considering the confirmed cases and estimating the undetected ones. In particular, the SIR or SEIR compartmental models have the advantage of providing useful insights for health care decision makers and of being easily manageable from a mathematical point of view.[2,19] Moreover, if adequately calibrated with good literature-based parameters, the simplicity of these models can avoid overfitting results.[2] The supposed proportion of undetected cases is reported to be highly variable. Its estimation from case fatality rates was already proposed,[1] but its integration with background death records data can further improve this estimate.

In the available published literature, searching PubMed between 01.12.2019 and 19.04.2020, several papers modelling the COVID-19 pandemic using SEIR models were found, most of them studied the evolution of the epidemic in Wuhan or Hubei or in single geographical context and proposed different scenarios depending on different transmission rates. However, only few of them took into account a measurable effect of public health intervention (e.g., mobility changes).[20] The work by Gatto et al.[20] is of major interest, as they apply a modified SEIR model (SE-

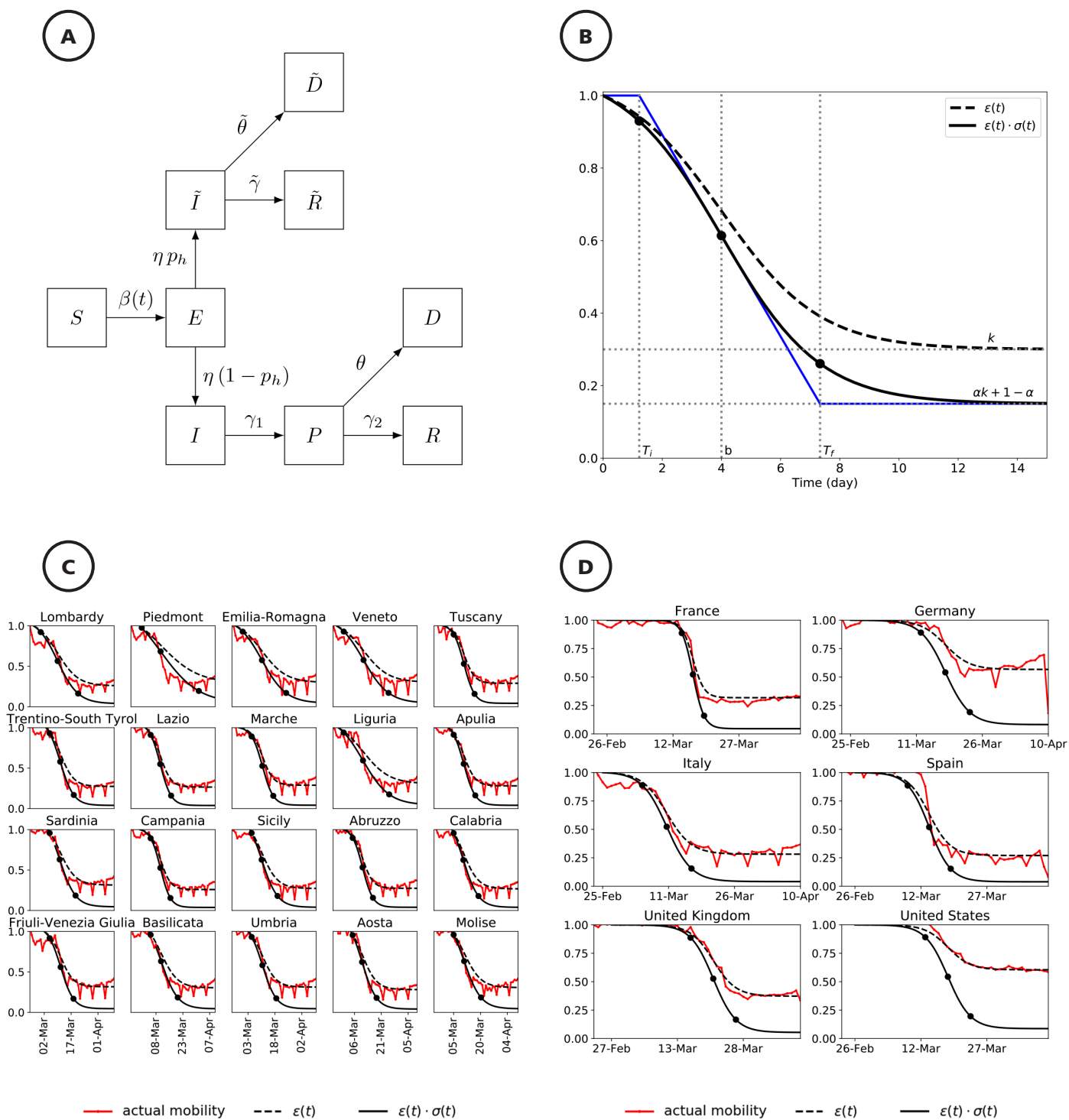

**Figure 1. A.** The scheme of the compartmental model: susceptible (S); exposed (E); infected (I); positive tested (P); recovered (R); deaths (D); ~ indicates undetected groups. **B.** The method for the ε(t)σ(t) logistic function building, extensively treated in supplementary methods considering both mobility and personal protective measures (PPM) effect: *k*, plateau level of mobility decreasing; ε(t), logistic function interpolating the mobility changes; α, coefficient representing the effect of PPM; Ti, start of public health measures; b, inflection time point of the logistic functions ε(t)σ(t) and ε(t). **C.** and **D.** showing the ε(t)σ(t) logistic functions (black thick lines) used in each scenario considering measured actual Google data of mobility changes, in Italian regions and across countries, respectively.
**Figura 1. A.** Lo schema del modello a compartimenti: suscettibili (S); esposti (E); infetti (I); testati positivi (P); guariti (R); morti (D); ~ indica i nascosti. **B.** La funzione logistica ε(t)σ(t) , discussa nei materiali supplementari, considera l'effetto sia della mobilità sia delle misure di protezione personale (PPM): *k*, plateau della mobilità; ε(t), funzione logistica che interpola I cambiamenti nella mobilità; α, coefficiente che rappresenta l'effetto delle PPM; Ti, inizio delle misure di sanità pubblica; b, punto di flesso delle funzioni logistiche ε(t)σ(t) e ε(t). **C.** e **D.** le funzioni logistiche ε(t)σ(t) (linea nera marcata) utilizzate in ogni scenario considerando i dati Google sulle variazioni nella mobilità, rispettivamente nelle regioni italiane e nei vari stati.

| REGION | No. (M) | $D_H/D_P$ | H (%) | $\beta(0)$ | $R(0)_P$ | $R(0)_H$ | $R(0)$ | $1/\gamma_2$ | MAPE (%) | ACTUAL POS. TEST. 24/4 *1,000 | EST. POS. TEST. 24/4 *1,000 (CI95%) | EST. TOT. INF. 24/4 *1,000 (CI95%) | POPOL. TOT. INF. 24/4 % (CI95%) | IH/P (%) | IH/T (%) |
|---|---|---|---|---|---|---|---|---|---|---|---|---|---|---|---|
| Abruzzo | 1.3 | 0.30 | 89 | 1.96 | 9.4 | 19.7 | 18.5 | 100 | 3.8 | 2.8 | 2.7 (2.0-4.4) | 26 (22-38) | 2.0 (1.7-2.9) | 93.3 | 92.9 |
| Valle d'Aosta | 0.1 | 0.00 | 89 | 1.77 | 8.5 | 17.8 | 18.5 | 33 | 12.3 | 1.1 | 1.3 (1.0-1.9) | 12 (11-15) | 9.6 (8.5-12.3) | 92.9 | 92.5 |
| Apulia | 4.0 | 1.40 | 93 | 2.07 | 9.9 | 20.8 | 20 | 100 | 2.8 | 3.9 | 4.0 (2.9-6.9) | 63 (51-95) | 1.6 (1.3-2.4) | 95.9 | 95.6 |
| Basilicata | 0.6 | 5.00 | 95 | 1.47 | 7 | 14.7 | 17.8 | 75 | 31.4 | 0.4 | 0.5 (0.4-0.7) | 11 (10-14) | 2.0 (1.8-2.5) | 97.1 | 97.0 |
| Calabria | 2.0 | 0.26 | 85 | 2.20 | 10.5 | 22 | 21.2 | 82 | 28.0 | 1.1 | 1.4 (1.1-2.3) | 10 (8-15) | 0.5 (0.4-0.7) | 90.3 | 89.9 |
| Campania | 5.8 | 0.00 | 79 | 1.79 | 8.6 | 17.9 | 16 | 100 | 3.2 | 4.3 | 4.1 (3.3-6.0) | 21 (19-26) | 0.4 (0.3-0.5) | 86.6 | 85.8 |
| Emilia-Romagna | 4.5 | 0.65 | 93 | 1.10 | 5.3 | 11 | 10.6 | 42 | 13.2 | 24.0 | 27.3 (23.1-35.5) | 426 (414-470) | 9.6 (9.3-10.5) | 95.9 | 95.7 |
| Friuli Venezia Giulia | 1.2 | 0.00 | 82 | 1.53 | 7.3 | 15.4 | 13.9 | 29 | 3.7 | 2.9 | 3.0 (2.4-4.5) | 17 (16-22) | 1.4 (1.3-1.8) | 88.4 | 87.7 |
| Lazio | 5.9 | 0.00 | 76 | 2.50 | 12 | 25.1 | 21.9 | 77 | 2.7 | 6.1 | 6.4 (4.7-11.4) | 29 (24-45) | 0.5 (0.4-0.8) | 84.2 | 83.5 |
| Liguria | 1.6 | 0.05 | 90 | 1.01 | 4.9 | 10.2 | 9.6 | 33 | 0.9 | 7.2 | 7.1 (6.1-9.1) | 77 (76-83) | 5 (4.9-5.4) | 94.1 | 93.8 |
| Lombardy | 10.1 | 0.58 | 95 | 0.97 | 4.6 | 9.7 | 9.5 | 32 | 8.6 | 71.3 | 63.6 (55.8-78.2) | 1,370 (1,353-1,395) | 13.6 (13.4-13.9) | 97.1 | 96.9 |
| Marche | 1.5 | 0.10 | 91 | 1.77 | 8.5 | 17.8 | 16.9 | 40 | 18.6 | 6.0 | 7.3 (5.5-11.3) | 82 (70-110) | 5.4 (4.6-7.2) | 94.2 | 93.9 |
| Molise | 0.3 | 1.84 | 91 | 0.90 | 4.3 | 9 | 8.9 | 100 | 7.9 | 0.3 | 0.3 (0.2-0.3) | 3.0 (3.0-3.1) | 1.0 (1.0-1.0) | 94.4 | 94.0 |
| Piedmont | 4.4 | 0.69 | 92 | 0.93 | 4.4 | 9.3 | 9.3 | 47 | 12.4 | 23.8 | 26.9 (23.9-32.2) | 373 (369-377) | 8.6 (8.5-8.7) | 95.1 | 94.9 |
| Sardinia | 1.6 | 4.65 | 96 | 1.62 | 7.8 | 16.2 | 19.9 | 57 | 29.7 | 1.3 | 1.7 (1.3-2.7) | 50 (43-68) | 3.1 (2.6-4.1) | 97.7 | 97.6 |
| Sicily | 5.0 | 0.49 | 90 | 1.53 | 7.3 | 15.3 | 16 | 98 | 12.4 | 3.0 | 3.3 (2.7-4.7) | 37 (34-45) | 0.7 (0.7-0.9) | 94.1 | 93.8 |
| Trentino-South Tyrol | 1.1 | 0.83 | 86 | 2.26 | 10.8 | 22.7 | 21 | 42 | 9.7 | 6.2 | 7.0 (5.1-11.7) | 53 (44-78) | 5.0 (4.1-7.3) | 91.1 | 90.7 |
| Tuscany | 3.7 | 0.00 | 89 | 1.74 | 8.4 | 17.5 | 16.5 | 100 | 2.2 | 8.9 | 9.2 (7-14.3) | 86 (74-116) | 2.3 (2.0-3.1) | 93.0 | 92.6 |
| Umbria | 0.9 | 1.23 | 84 | 1.48 | 7.1 | 14.9 | 13.6 | 23 | 34.8 | 1.4 | 1.9 (1.5-2.8) | 13 (12-16) | 1.5 (1.3-1.8) | 90.2 | 89.6 |
| Veneto | 4.9 | 0.57 | 85 | 0.91 | 4.4 | 9.1 | 8.4 | 61 | 5.3 | 17.2 | 17.9 (16-21.6) | 127 (125-128) | 2.6 (2.5-2.6) | 90.7 | 90.2 |
| France | 67.0 | 0.53 | 91 | 1.71 | 8.2 | 17.1 | 16.3 | 66 | 31.8 | 158.6 | 110.7 (84.5-170.6) | 1,376 (1,186-1,852) | 2.1 (1.8-2.8) | 94.6 | 94.3 |
| Germany | 83.0 | 0.53 | 60 | 1.72 | 8.3 | 17.3 | 13.6 | 15 | 30.0 | 155.0 | 214.0 (159.8-396.4) | 610 (501-1,020) | 0.7 (0.6-1.2) | 71.5 | 70.8 |
| Italy | 60.3 | 0.53 | 93 | 1.17 | 5.6 | 11.7 | 11.2 | 81 | 18.8 | 193.0 | 153.0 (129.2-202.4) | 2,212 (2,149-2,476) | 3.7 (3.6-4.1) | 95.6 | 95.3 |
| Spain | 46.9 | 0.53 | 91 | 1.92 | 9.2 | 19.3 | 18.4 | 26 | 10.2 | 219.8 | 193.1 (148.8-295.6) | 2,294 (2,003-3,048) | 4.9 (4.3-6.5) | 94.4 | 94.1 |
| United Kingdom | 66.7 | 0.53 | 92 | 1.27 | 6.1 | 12.7 | 12.2 | 10,000† | 24.9 | 143.5 | 102.4 (84.3-139.5) | 1,474 (1,374-1,739) | 2.2 (2.1-2.6) | 95.3 | 95.0 |
| United States | 328 | 0.53 | 76 | 1.38 | 6.6 | 13.9 | 12.1 | 100 | 8.5 | 905.4 | 814 (574-1,415) | 3,785 (2,949-5,911) | 1.2 (0.9-1.8) | 84.0 | 83.5 |

**No.:** population size / popolazione; **M:** millions of inhabitants / milioni di abitanti; $D_H/D_P$: ratio between estimated undetected COVID-19 deaths and reported ones from Italian death records data, countries values are assumed equal to the Italian one / rapporto tra la stima di morti nascoste dovute al COVID-19 e quelle riportate dai registri di mortalità italiani, per gli altri stati assumiamo che i valori siano simili a quelli dell'Italia; **H:** percentage fraction of undetected cases / frazione di casi nascosti; $\beta(0)$: transmission rate at time 0 / tasso di trasmissione al tempo 0; $R(0)_P$: basic reproductive number imputable to positive tested cases / numero di riproduzione di base imputabile ai casi positivi testati; $R(0)_H$: basic reproductive number imputable to undetected cases / numero di riproduzione di base imputabile ai casi nascosti; $R(0)$: overall basic reproductive number / numero di riproduzione di base totale; $1/\gamma_2$: fitted recovery notification time / tempo di comunicazione della guarigione (fittato); **MAPE:** mean absolute percentage error / errore assoluto percentuale medio; **CI:** reasonable confidence interval / intervallo di confidenza ragionevole; **IH/P:** fraction of detected cases caused by undetected ones until 24.04.2020 / frazione dei casi diagnosticati generati da casi non diagnosticati fino al 24.04.2020; **IH/T:** fraction of the overall new cases caused by undetected ones until 24.04.2020 / frazione dei nuovi casi totali generati da casi non diagnosticati fino al 24.04.2020; †since 13.04.2020, public data regarding reported recovery cases are not available anymore / dopo il 13.04.2020, i dati riguardanti il numero di guariti non sono stati pubblicati

**Table 1.** Results of the SEIR model applied in several different Regions.
**Tabella 1.** Risultati del modello SEIR applicati in diverse regioni.

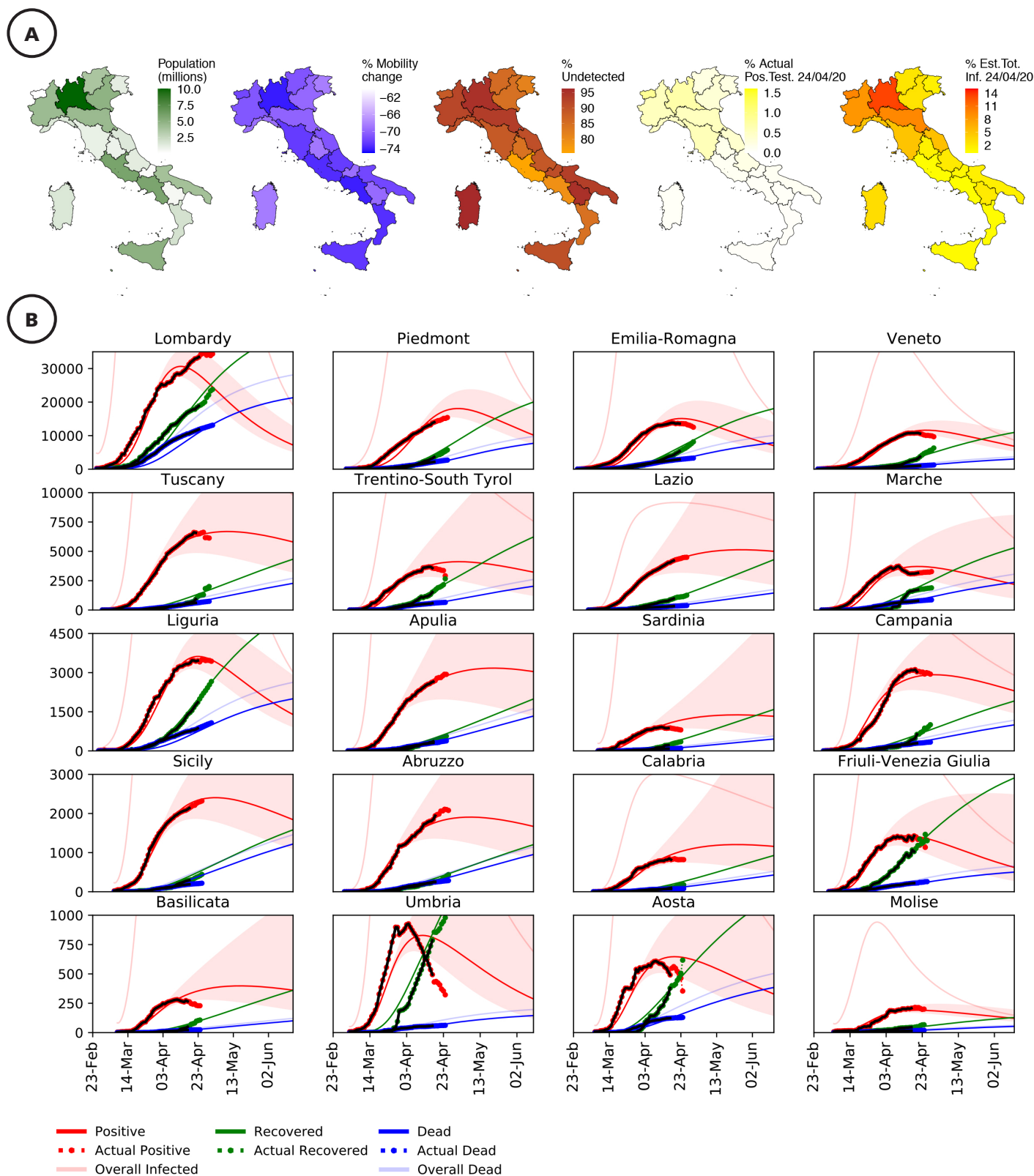

**Figure 2. A.** Heatmaps of the Italian Regions illustrating: heterogeneity across region of population size; mean mobility change (11th-17th April 2020) compared to baseline; estimated percentage fraction of undetected COVID-19 cases; actual cumulative fraction of population positive tested, and estimated cumulative fraction of overall infected cases (including undetected) until 24.04.2020. **B.** Epidemic predicted and actual reported data in different Italian Regions; overall infected and deaths include undetected cases. The models were fitted on actual data until 17 April (black line), cumulative observed values from 18 April to 24 April are used for judging models accuracy. The shadow regions represent the reasonable confidence intervals.

**Figura 2. A.** Mappe coropletiche delle regioni italiane che mostrano: l'eterogeneità interregionale del numero di abitanti, il cambiamento medio della mobilità (11-17 aprile 2020) comparato al livello basale; la frazione stimata di casi nascosti di COVID-19; la frazione effettiva (cumulata) della popolazione testata positiva e la stima di tutti gli infetti (compresi i nascosti) fino al 24.04.2020. **B.** I dati dell'epidemia stimati ed effettivi riportati in diverse regioni italiane; gli infetti totali e le morti, compresi i casi nascosti. Il modello è stato fittato utilizzando i dati fino al 17 aprile (linea nera), i valori cumulati osservati dal 18 aprile al 24 aprile sono utilizzati per valutare l'accuratezza del modello. Le aree ombreggiate rappresentano l'intervallo di confidenza ragionevole.

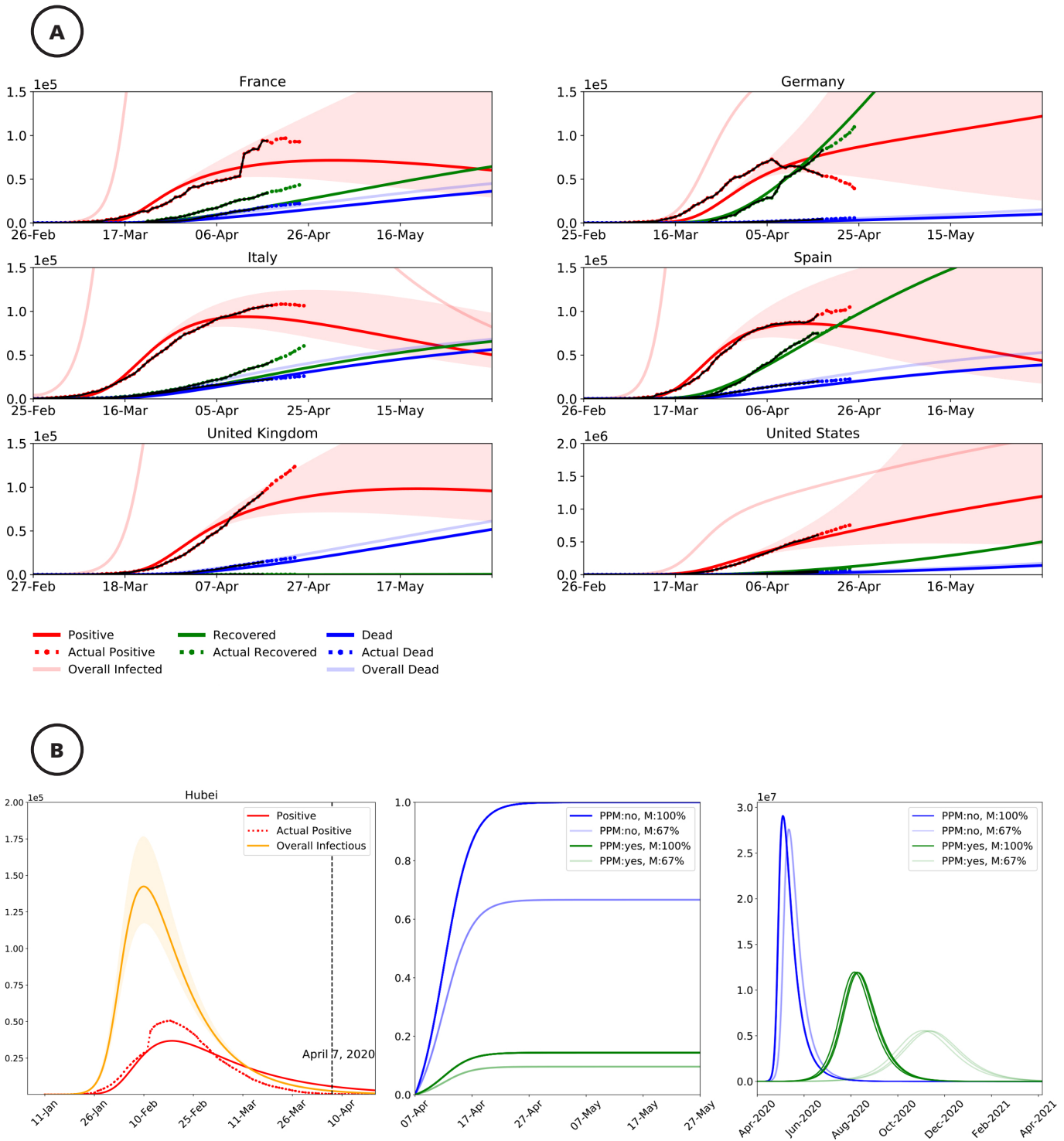

**Figure 3. A.** Epidemic predicted and actual reported data in different countries; overall infected and deaths include undetected cases. The models were fitted on actual data until 17 April (black line), cumulative observed values from 18 April to 24 April are used for judging the accuracy of the models. The shadow regions represent the reasonable confidence intervals. **B.** From left to right: the left panel shows the estimated positive detected fitted cases in Hubei region, infectious patients (undetected or not already detected cases) are represented with orange line with shadow showing the reasonable confidence of interval; the central panel shows the shape of different $\varepsilon(t)\sigma(t)$ functions after the restoration of partial or full mobility (M), with or without the use of personal protective measures (PPM) after 7 April; the right panel shows different scenarios of overall infected secondary epidemic peaks adopting different strategies of lockdown's mitigation.
**Figura 3. A.** I dati dell'epidemia previsti e quelli effettivi riportati in diversi paesi; gli infetti totali e le morti includono i casi nascosti. Il modello è stato fittato utilizzando i dati disponibili fino al 17 Aprile (linea nera), i valori cumulati osservati dal 18 Aprile al 24 Aprile sono utilizzati per valutare l'accuratezza del modello. Le aree ombreggiate rappresentano l'intervallo di confidenza ragionevole. **B.** Da sinistra a destra: il pannello di sinistra mostra la stima dei casi positivi testati fittata nella regione di Hubei, i pazienti infettivi (nascosti o non ancora testati) sono rappresentati da una linea arancione con ombreggiatura che mostra l'intervallo di confidenza ragionevole; il pannello centrale mostra la forma delle funzioni $\varepsilon(t)\sigma(t)$ dopo un parziale o totale ripristino della mobilità (M), con o senza l'utilizzo di misure di protezione personale (PPM) dopo il 7 Aprile; il pannello di destra mostra diversi scenari di picchi epidemici secondari utilizzando diverse strategie per l'allentamento del lockdown.

PIA) using data of geolocated smartphone users[21] to estimate the effect of lockdown intervention measures on the Italian epidemic evolution.

The results of this study in the Italian case-study report the cumulative number of overall affected cases as of 25 March and 28 March equal to $1.46*10^6$ (95%CI $1.31*10^6$-$1.60*10^6$) and $1.60*10^6$ (95%CI $1.48*10^6$-$1.76*10^6$), respectively. The estimates here presented have the same order of magnitude of those by Gatto et al. (median $0.6*10^6$ on 25 March) and those by the Imperial college group[1] ($5.9*10^6$ on 28 March; 95%CI $1.9*10^6$-$15.2*10^6$), much higher than the 74,286 and 92,472 confirmed cases until these dates.

To the best of knowledge of the Authors of this papers, this is the first SEIR compartmental model taking into account the mobility changes measured by Google in order to track the effect of lockdown measures so far. The decreasing logistic function permitted to analytically solve the system and to integrate the effect of the adoption of personal protective measures (hand washing an masks wearing) derived from literature estimates.[11]

Fitting the recovery reporting time after the isolation of positive tested cases permitted to depict wide differences among countries (IQR 36-95 days), but also in-between Italian Regions (IQR 38-99 days); of note, the fitted recovery reporting time of 10,000 days (the maximum of the permitted range of values) for UK is the consequence of the interruption of disclosing publicly recovery data since the beginning of April 2020. Recovery reporting time heterogeneity is of main interest as the removal time of recovered individuals is expected to influence the shape of the epidemic curves of active cases before and after relaxing the lockdown measures. Furthermore, as shown in the Hubei case-study (figure 3B), the outputs of the model presented in this study in term of infectious groups (undetected or not already confirmed cases) can be used to draw possible different trajectories of secondary epidemics after the abolishment or mitigation of mobility restrictions, considering different kind of self-protective behaviour.

Even though specific and quantitative predictions from a mathematical model need to be taken with caution, some key messages can be derived, as the estimation and the evolution of complex phenomena that are hardly imaginable by the common sense, for instance the hidden dynamic of undetected cases parallel to the dynamic of detected ones.[22] Despite having some limitations, not considering spatio-temporal dynamics and stochastic features, the main advantage of this system is its easy application, that can be even improved by further updating mobility information, nowadays publicly available, and acquiring Region-specific background death records data.

## REFERENCES

1. Flaxman S, Mishra S, Gandy A, et al. Estimating the number of infections and the impact of non-pharmaceutical interventions on Covid-19 in 11 European countries. London: Imperial College London; 2020. Available from: https://doi.org/10.25561/77731
2. Petard H. A contribution to the mathematical theory of big game hunting. Am Math Mon 1938;45(7):446-47.
3. Dong E, Du H, Gardner L. An interactive web-based dashboard to track Covid-19 in real time. Lancet Infect Dis 2020;20(5):533-34.
4. Lauer SA, Grantz KH, Bi Q, et al. The incubation period of coronavirus disease 2019 (Covid-19) from publicly reported confirmed cases: estimation and application. Ann Intern Med 2020;172(9):577-82.
5. Kraemer MUG, Yang CH, Gutierrez B, et al. The effect of human mobility and control measures on the Covid-19 epidemic in China. Science 2020;368(6490):493-97.
6. Wang Y, Liu Y, Liu L, Wang X, Luo N, Li L. Clinical outcomes in 55 patients with severe acute respiratory syndrome coronavirus 2 who were asymptomatic at hospital admission in Shenzhen, China. J Infect Dis 2020;221(11):1770-74.
7. Verity R, Okell LC, Dorigatti I, et al. Estimates of the severity of coronavirus disease 2019: a model-based analysis. Lancet Infect Dis 2020;20(6):669-77.
8. Shim E, Mizumoto K, Choi W, Chowell G. Estimating the risk of Covid-19 death during the course of the outbreak in Korea, February-May 2020. J Clin Med 2020;9(6):1641.
9. Citato come report 88 ma viene linkato il n. 87. World Health Organisation. Coronavirus disease 2019 (Covid-19). Situation Report – 88. Geneva: WHO; 2020. Available from: https://www.who.int/docs/default-source/coronaviruse/situation-reports/20200416-sitrep-87-covid-19.pdf?sfvrsn=9523115a_2
10. World Health Organisation. Coronavirus disease (Covid-19) advice for the public. Geneva: WHO; 2020. Available from: https://www.who.int/emergencies/diseases/novel-coronavirus-2019/advice-for-public (last accessed: 29.04.2020)
11. Jefferson T, Foxlee R, Del Mar C, et al. Physical interventions to interrupt or reduce the spread of respiratory viruses: systematic review. BMJ 2008;336(7635):77-80.
12. Lewis CD. Industrial and business forecasting methods: a practical guide to exponential smoothing and curve fitting. Boston: Butterworth scientific; 1982.
13. Signorelli C, Scognamiglio T, Odone A. Covid-19 in Italy: impact of containment measures and prevalence estimates of infection in the general population. Acta Biomed 2020;91(3-S):175-79.
14. Van Den Driessche P, Watmough J. Reproduction numbers and sub-threshold endemic equilibria for compartmental models of disease transmission. Math Biosci 2002;180(1-2):29-48.
15. De Simone A, Piangerelli M. The impact of undetected cases on tracking epidemics: the case of Covid-19. arXiv 2020;200506180. doi: 10.1016/j.chaos.2020.110167
16. Li R, Pei S, Chen B, et al. Substantial undocumented infection facilitates the rapid dissemination of novel coronavirus (SARS-CoV2). Science 2020;368(6490):489-93
17. Wu Z, McGoogan JM. Characteristics of and Important lessons from the coronavirus disease 2019 (Covid-19) outbreak in China: Summary of a report of 72 314 cases from the chinese Center for Disease control and prevention. JAMA 2020;323(13):1239-42.
18. Wilder-Smith A, Freedman DO. Isolation, quarantine, social distancing and community containment: pivotal role for old-style public health measures in the novel coronavirus (2019-nCoV) outbreak. J Travel Med 2020;27(2)taaa020.
19. Blackwood JC, Childs LM. An introduction to compartmental modeling for the budding infectious disease modeler. Lett Biomath 2018;5(1):195-221.
20. Gatto M, Bertuzzo E, Mari L, et al. Spread and dynamics of the Covid-19 epidemic in Italy: Effects of emergency containment measures. Proc Natl Acad Sci USA 2020;117(19):10484-91
21. Pepe E, Bajardi P, Gauvin L, et al. Covid-19 outbreak response: a first assessment of mobility changes in Italy following national lockdown. medRxiv 2020;2020.03.22.20039933. Available from: http://medrxiv.org/content/early/2020/04/07/2020.03.22.20039933
22. Jewell NP, Lewnard JA, Jewell BL. Predictive mathematical models of the Covid-19 pandemic: underlying principles and value of projections. JAMA 2020;323(19):1893-94.